\documentclass[prb,twocolumn,superscriptaddress,showpacs,amsmath,amssymb]{revtex4-1}

\usepackage{graphicx}% Include figure files
\usepackage{dcolumn}% Align table columns on decimal point
\usepackage{bm}% bold math
\usepackage{epstopdf}
\usepackage{upgreek}

\begin{document}

\title{Advanced first-principles theory of superconductivity including both lattice vibrations and spin fluctuations: the case of FeB$_4$}

\author{J. Bekaert}
\email{jonas.bekaert@uantwerpen.be}
\affiliation{%
 Department of Physics, University of Antwerp,
 Groenenborgerlaan 171, B-2020 Antwerp, Belgium
}%
 \author{A. Aperis}%
 \email{alex.aperis@physics.uu.se}
 \affiliation{%
 Department of Physics and Astronomy, 
 Uppsala University, Box 516, 
 SE-751 20 Uppsala, Sweden
}%
 \author{B. Partoens}
 \affiliation{%
 Department of Physics, University of Antwerp,
 Groenenborgerlaan 171, B-2020 Antwerp, Belgium
}%
\author{P. M. Oppeneer}
 \affiliation{%
Department of Physics and Astronomy, Uppsala University,
Box 516, SE-751 20 Uppsala, Sweden
}
 \author{M. V. Milo\v{s}evi\'{c}}
\email{milorad.milosevic@uantwerpen.be}
 \affiliation{%
 Department of Physics, University of Antwerp,
 Groenenborgerlaan 171, B-2020 Antwerp, Belgium
}%

\date{\today}

\begin{abstract}
\noindent  We present an advanced method to study spin fluctuations in superconductors quantitatively, and entirely from first principles. 
This method can be generally applied to materials where electron-phonon coupling and spin fluctuations coexist. We employ it here to examine the recently synthesized superconductor iron tetraboride (FeB$_4$) with experimental $T_{\mathrm{c}}\sim 2.4$ K [H. Gou \textit{et al.}, Phys. Rev. Lett. \textbf{111}, 157002 (2013)]. We prove that FeB$_4$ is particularly prone to ferromagnetic spin fluctuations due to the presence of iron, resulting in a large Stoner interaction strength, $I=1.5$ eV, as calculated from first principles. The other important factor is its Fermi surface that consists of three separate sheets, among which two nested ellipsoids. The resulting susceptibility has a ferromagnetic peak around $\textbf{q}=0$, from which we calculated the repulsive interaction between Cooper pair electrons using the random phase approximation. Subsequently, we combined the electron-phonon interaction calculated from first principles with the spin fluctuation interaction in fully anisotropic Eliashberg theory calculations. We show that the resulting superconducting gap spectrum is conventional, yet very strongly depleted due to coupling to the spin fluctuations. The critical temperature decreases from $T_{\mathrm{c}}= 41$ K, if they are not taken into account, to $T_{\mathrm{c}}= 1.7$ K, in good agreement with the experimental value.

\end{abstract}

\maketitle

\section{Introduction}

Spin fluctuations are magnetic excitations in materials without long-range magnetic order. Ferromagnetic spin fluctuations (FSFs), or paramagnons, specifically arise in materials which are close to ferromagnetic instabilities, as described by Stoner theory \cite{Moriya2006}. In the case of spin singlet Cooper pairing, the interaction between electrons mediated via FSFs is repulsive, and therefore competing with Cooper pairing, in addition to the Coulomb interaction between electrons \cite{Scalapino1999}. On the other hand, the opposite is true for spin triplet pairing, where paramagnons are considered as the primary mediators \cite{Scalapino1999}, although not the only ones \cite{Aperis2011}.

Competition between attractive electron-phonon interaction and the repulsive interaction mediated by FSFs forms a long standing problem that emerged less than a decade after the theory of Bardeen-Cooper-Schrieffer (BCS) \cite{PhysRevLett.17.433}. Among the earliest attempts for a quantitative analysis, Riblet introduced the coupling to FSFs in the isotropic McMillan formula for the critical temperature ($T_{\mathrm{c}}$) \cite{PhysRevB.3.91,Grimvall}. Similar attempts have also been made for antiferromagnetic spin fluctuations \cite{PhysRevB.45.13047}. Dolgov \textit{et al.} subsequently derived an improved McMillan formula for $T_{\mathrm{c}}$ \cite{PhysRevLett.95.257003}. The latter can be combined with first-principles calculations as was done for, e.g., hole-doped CuBiSO, where pairing to spin fluctuations was found to be very strong and able to induce spin triplet superconductivity under certain doping conditions \cite{PhysRevB.83.100505}. Another notable example where spin triplet superconductivity is rather well established and where FSFs have been proposed to play a role is Sr$_2$RuO$_4$ ($T_{\mathrm{c}}=1.5$ K) \cite{PhysRevLett.79.733}, although the microscopic pairing mechanism is still not completely understood (see Refs.~\citenum{RevModPhys.75.657} and \citenum{RevMaeno} for reviews on this topic).   

We revisit here the question of spin fluctuations, with a new and advanced computational method. It consists of first calculating the microscopic pairing mechanisms, i.e., electron-phonon coupling and coupling of electrons to FSFs. The electron-phonon interaction is calculated using density functional perturbation theory (DFPT) \cite{PhysRevB.54.16487}, similar to what is done in, e.g., Refs.~\citenum{PhysRevB.92.054516,PhysRevB.94.144506,Bekaert2017,Bekaert2017b}. We treat spin fluctuations by means of the random phase approximation (RPA), afterwards building it into the anisotropic Eliashberg equations. Specifically, we calculate the susceptibility from the electronic band structure as well as the interaction strength, in this case the Stoner interaction strength. Subsequently, we self-consistently solve the multiband anisotropic Eliashberg equations with the full \textit{ab initio} calculated input \cite{Choi2002, PhysRevB.87.024505, PhysRevB.92.054516,PhysRevB.94.144506,PhysRevB.94.144506,Bekaert2017,Bekaert2017b}.

We applied this technique successfully to the recently discovered superconductor iron tetraboride (FeB$_4$). A famous example of first-principles materials design, superconductivity in FeB$_4$ was first predicted \emph{in silico} by Kolmogorov \textit{et al.} in 2010 \cite{PhysRevLett.105.217003}, after which the material was synthesized and measured to be superconducting with $T_{\mathrm{c}}\sim 2.4$ K by Gou \textit{et al.} in 2013 \cite{PhysRevLett.111.157002}. The crystal structure of FeB$_4$ is orthorhombic and consists of FeB$_{12}$ polyhedra stacked in columns along the $\textbf{a}$-direction (where we defined $a < b < c$). A more detailed description of the crystal structure can be found in Appendix A. This crystal structure, and in particular the presence of the light element boron, gives FeB$_4$ a very high mechanical hardness \cite{doi:10.1063/1.3556564,PhysRevLett.111.157002,doi:10.1063/1.4871627,Kotmool2014}. 

Our motivation to study FeB$_4$ in more depth derives from several aspects. First of all, its $T_{\mathrm{c}}$ was severely overestimated (by an order of magnitude) in the theory of Ref.~\citenum{PhysRevLett.105.217003} with respect to the experimental value \cite{PhysRevLett.111.157002}. The prediction was based on the isotropic McMillan-Dynes formula where only electron-phonon interaction and usual Coulomb repulsion was taken into account. We recognized this as a smoking gun for unconventional interactions in FeB$_4$, which we prove to be FSFs in this paper. Secondly, superconductors containing Fe have attracted much interest recently, since the discovery of superconductivity in the iron-pnictides (e.g., F-doped LaFeAsO \cite{doi:10.1021/ja800073m}), the iron-arsenides (e.g., Ba$_{1-x}$K$_x$Fe$_2$As$_2$ \cite{PhysRevLett.101.107006}, and LiFeAs \cite{Wang2008538}), and the iron-chalcogenides (e.g., FeSe \cite{Hsu23092008,doi:10.1063/1.3000616}). In this respect, our analysis contributes to the understanding of the microscopic mechanisms at work in the Fe-based superconductors. We must stress, however, that the properties of spin fluctuations in FeB$_4$ are fundamentally different from those of other Fe-based superconductors, e.g., the ferro-pnictides. F-doped LaFeAsO, for instance, shows a susceptibility peak at nonzero $\textbf{q}=\left(\frac{\pi}{a},\frac{\pi}{a}\right)$, as discovered by Mazin \textit{et al.} \cite{PhysRevLett.101.057003}, and thus a tendency for antiferromagnetic spin fluctuations. As such, our study establishes Fe-based superconductors as a diverse family, in which various different types of spin fluctuations occur. Thirdly, the multiband and multigap superconductivity in borides such as MgB$_2$ \cite{Nagamatsu2001,0953-2048-16-2-305,Choi2002,PhysRevB.91.214519,PhysRevB.87.024505,
PhysRevB.92.054516,Bekaert2017,Bekaert2017b}, OsB$_2$ \cite{PhysRevB.82.144532,PhysRevB.94.144506}, and ZrB$_{12}$ \cite{PhysRevB.72.024547,PhysRevB.72.024548} is known to be very rich, and consequently possible relations to superconductivity in FeB$_4$ are worthy of further exploration.

This paper is organized as follows. First, in Sec.~II, we elaborate on the methodology we develop in this work, building FSFs calculated from first-principles, into the anisotropic Eliashberg equations. In Sec.~III, we discuss the multiband electronic structure of FeB$_4$ (showing for the first time its Fermi surface) and the electron-phonon (\textit{e-ph}) interaction in this compound. We proceed by our first-principles calculations of the FSFs and their coupling to the electrons in FeB$_4$ in Sec.~IV. This is followed by a discussion of the superconducting properties of FeB$_4$, the gap spectrum and the very good agreement between the theoretical and experimental $T_{\mathrm{c}}$, in Sec.~V. Finally, our conclusions are given in Sec.~VI.

\section{Methodology}
\label{Sec2}

First, we will establish how FSFs can be built into the anisotropic Eliashberg equations within the random phase approximation (RPA). The tendency for spin fluctuations is mainly determined by the susceptibility, and in particular its behavior at the Fermi level ($E_{\mathrm{F}}$). The bare (i.e., noninteracting) susceptibility at $E_{\mathrm{F}}$ (known as the Lindhard function), is given by the following function of momentum (\textbf{q}) and Matsubara frequencies ($\omega_n$)
\begin{align}
&\chi^0\left(\textbf{q},i\omega_n\right)=\sum_{jj'}\chi^0_{jj'}\left(\textbf{q},i\omega_n\right)=\nonumber \\
&\sum_{jj'}\sum_{\textbf{k}}\frac{n_{\mathrm{F}}\left(\xi_{\textbf{k},j}\right)-n_{\mathrm{F}}\left(\xi_{\textbf{k}+\textbf{q},j'}\right)}{\xi_{\textbf{k},j}-\xi_{\textbf{k}+\textbf{q},j'}+i\omega_n} \delta\left(\xi_{\textbf{k},j}\right)\delta\left(\xi_{\textbf{k}+\textbf{q},j'}\right)~,
\label{eq:bare_susc}
\end{align}
where $n_{\mathrm{F}}\left(\xi_{\textbf{k},j}\right)$ is the Fermi-Dirac distribution, $\xi_{\textbf{k},j}=E_{\textbf{k},j}-E_{\mathrm{F}}$ is the electronic band structure relative to $E_{\mathrm{F}}$, and where we sum over the electronic band indices $j$ and $j'$. To arrive at this expression, the constant matrix element approximation (CMEA) has been employed \cite{CohenLouie}. The Dirac $\delta$-functions are introduced in order to restrict the susceptibilities to the Fermi surface contributions. We evaluate $\delta\left(\xi_{\textbf{k},j}\right)$ numerically as $\delta\left(\xi_{\textbf{k},j}\right)=\frac{1}{\sqrt{\pi}\sigma}\mathrm{exp}\left(-\left(\frac{\xi_{\textbf{k},j}}{\sigma}\right)^2\right)$ with broadening $\sigma=0.01$ Ha. 

In compounds with more than one atomic species one needs to take into account that not necessarily all the electronic states are involved in the FSFs. This is only the case for the states belonging to the element(s) with a ferromagnetic tendency. To take the example of FeB$_4$, as we will show in Sec.~III, Fe lies at the origin of the fluctuations. This means that the susceptibility to FSFs needs to be normalized with the ratio of Fe-electronic states ($N_{\mathrm{Fe}}$) to the total intraband susceptibility in the limit $\textbf{q} \rightarrow 0$, $\omega \rightarrow 0$, i.e., $\sum_j \chi^0_{jj}(0,0)$. We denote this fraction as $\mathcal{F}_{\mathrm{Fe}}=N_{\mathrm{Fe}}/\sum_j \chi^0_{jj}(0,0)$. Thus, for the total susceptibility we can use the RPA expression
\begin{align}
\chi^{\mathrm{RPA}}\left(\textbf{q},i\omega_n\right)=\frac{\mathcal{F}_{\mathrm{Fe}}\chi^0\left(\textbf{q},i\omega_n\right)}{1-I\mathcal{F}_{\mathrm{Fe}}\chi^0\left(\textbf{q},i\omega_n\right)}~,
\label{eq:susc_RPA}
\end{align}
where $I$ is the ferromagnetic interaction strength. We will expand on how the latter can be calculated from first-principles in Sec.~IV. 

Based on the RPA susceptibility we can calculate the coupling of electrons to FSFs as
\begin{align}
\lambda_{\mathrm{sf}}\left(\textbf{q},i\omega_n\right)=\frac{3}{2}N_{\mathrm{Fe}}I^2\chi^{\mathrm{RPA}}\left(\textbf{q},i\omega_n\right)~.
\label{eq:coupling}
\end{align}

Finally, we include FSFs in the anisotropic Eliashberg equations \cite{PhysRevB.92.054516,PhysRevB.94.144506,Bekaert2017,Bekaert2017b} within spin singlet pairing by means of two pairing kernels, one expressing mass enhancement of the electrons ($K^+$), the other expressing the net coupling strength ($K^-$). As the electron mass is enhanced by both \textit{e-ph} interaction, $\lambda_{\mathrm{ep}}\left(\textbf{q},i\omega_n \right)$, and FSFs, the kernel $K^+\left(\textbf{q},i\omega_n \right)=\lambda_{\mathrm{ep}}\left(\textbf{q},i\omega_n \right)+\lambda_{\mathrm{sf}}\left(\textbf{q},i\omega_n \right)$. On the other hand, the coupling strength in the spin singlet case is depleted, as expressed by $K^-\left(\textbf{q},i\omega_n \right)=\lambda_{\mathrm{ep}}\left(\textbf{q},i\omega_n \right)-\lambda_{\mathrm{sf}}\left(\textbf{q},i\omega_n \right)$. The momentum-dependent \textit{e-ph} coupling, $\lambda_{\mathrm{ep}}\left(\textbf{q},i\omega_n \right)$, can be calculated within density functional perturbation theory (DFPT) \cite{PhysRevB.54.16487}. 

\begin{figure}[b]
\centering
\includegraphics[width=1\linewidth]{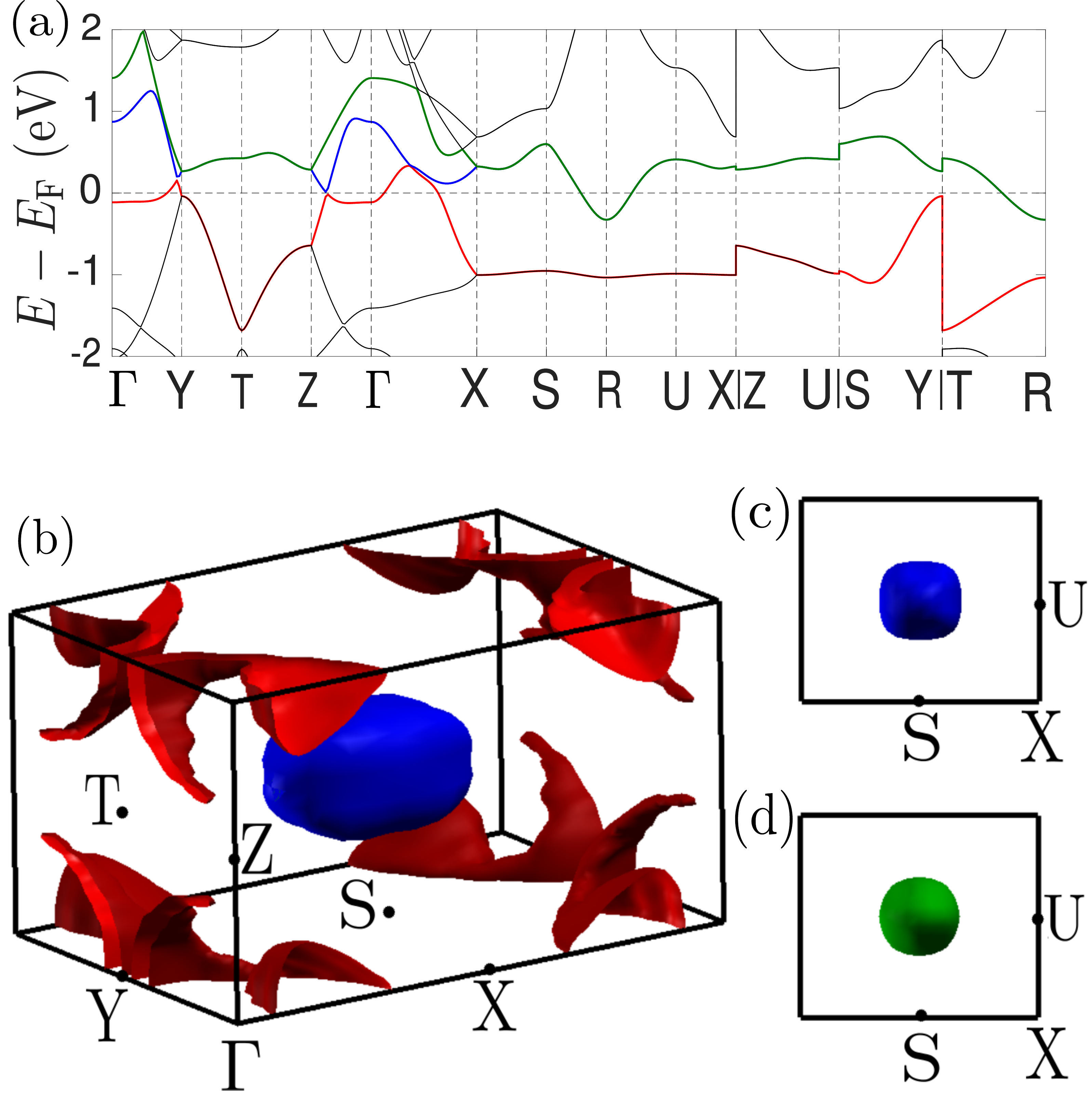}
\caption{(Color online) The electronic structure of FeB$_4$, calculated using DFT. (a) The band structure around the Fermi level ($E_{\mathrm{F}}$), where three bands are seen to cross $E_{\mathrm{F}}$. (b) The calculated Fermi surface of FeB$_4$, where the colors correspond to those of (a). It consists of two nested ellipsoids around high-symmetry point R (blue and green), as well as a third, anisotropic sheet (red). (c) and (d) The two nested ellipsoids pictured individually in frontal view.}
\label{fig:fig1}
\end{figure}
\begin{figure}[t]
\centering
\includegraphics[width=1\linewidth]{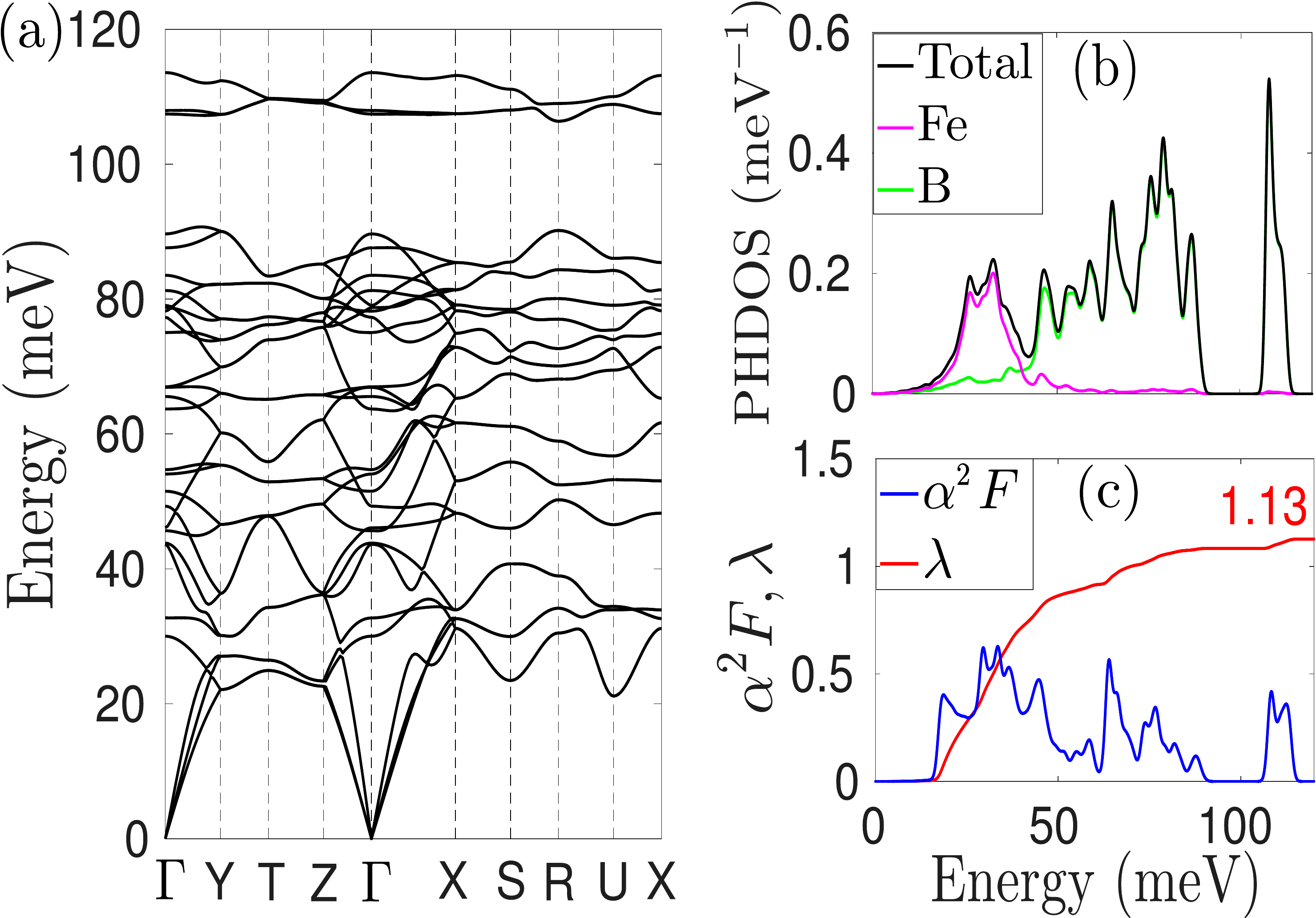}
\caption{(Color online) Phonons and electron-phonon coupling, calculated within DFPT. (a) The phonon dispersion, extending as high as $\sim 115$ meV. (b) The phonon density of states (PHDOS), including the contributions of Fe (purple) and B (green). (c) The isotropic Eliashberg function, obtained as $\alpha^2F(\omega)=\langle\langle\alpha^2F({\bf k\, k'},\omega)\rangle_{{\bf k}'_{\mathrm{F}}}\rangle_{{\bf k}_{\mathrm{F}}}$ (i.e., the double Fermi surface average of the full Eliashberg function), and the corresponding total \textit{e-ph} coupling.}
\label{fig:fig2}
\end{figure}
\begin{figure*}[]
\centering
\includegraphics[width=1\linewidth]{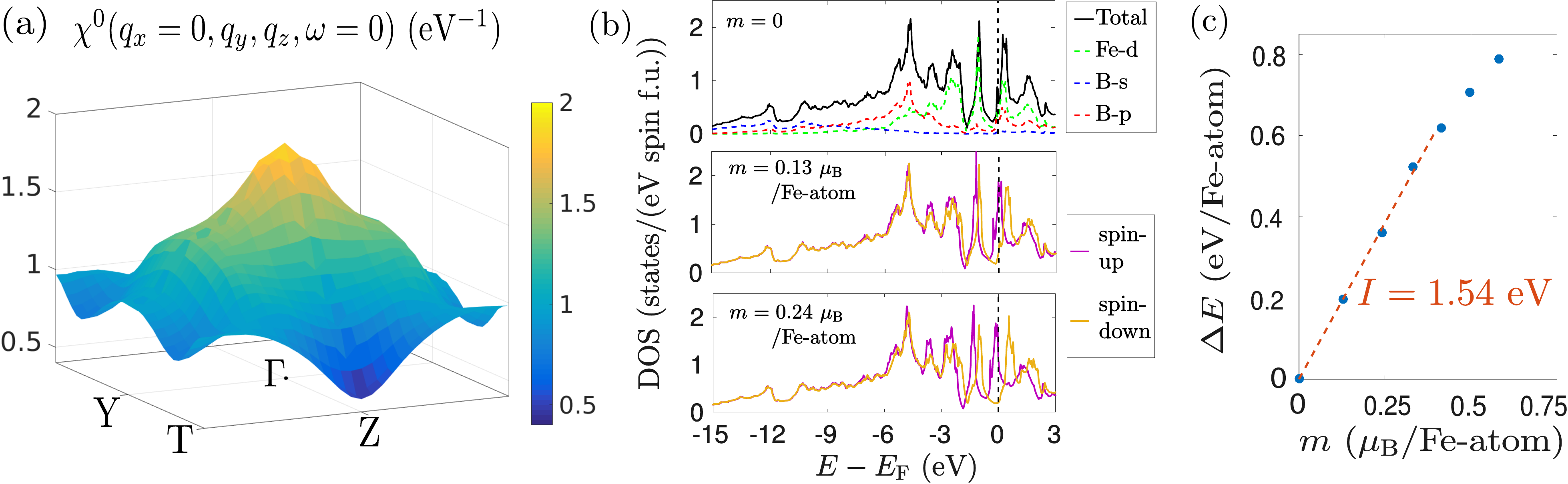}
\caption{(Color online) (a) The bare susceptibility of FeB$_4$, $\chi^0\left(\textbf{q},\omega \right)$, for $q_x=0$ and $\omega=0$ (static), calculated from the band structure at $T=1.5$ K. It shows a strong peak around the $\Gamma$ point, corresponding to FSFs. (b) The electronic density of states (DOS) per formula unit (f.u.) of FeB$_4$ for different levels of magnetization of the Fe atoms: the paramagnetic case with $m=0$ (top panel), $m=0.13~\mu_{\mathrm{B}}$/Fe-atom (middle panel), and $m=0.24~\mu_{\mathrm{B}}$/Fe-atom (lower panel), where in the latter two cases an energy shift between spin-up and spin-down states is apparent. In the top panel, the contributions of Fe-d, B-s and B-p states are also shown. (c) The corresponding energy shift, which obeys $\Delta E=I m / \mu_{\mathrm{B}}$ for small $m$ (dashed line), from which we obtain the Stoner parameter $I=1.54$ eV for FeB$_4$.}
\label{fig:fig3}
\end{figure*}

\section{Electronic structure and electron-phonon interaction}

Here, we apply the approach described in Sec. II to FeB$_4$, in order to demonstrate that quantitative results can be obtained. Our investigation starts from the electronic structure of FeB$_4$, near $E_{\mathrm{F}}$, calculated using density functional theory (DFT) as implemented in ABINIT \cite{Gonze20092582}. More detailed information on the first-principles calculations is given in Appendix B. The band structure of FeB$_4$ around $E_{\mathrm{F}}$ is shown in Fig.~\ref{fig:fig1}(a). Three bands are observed to cross $E_{\mathrm{F}}$, as indicated with three different colors (red, blue and green). The corresponding Fermi surface is displayed in Fig.~\ref{fig:fig1}(b), where the same colors are used. Around point R, the center of the cell, there are two nested ellipsoidal sheets (blue and green), while the third sheet (red) is more anisotropic. In Figs.~\ref{fig:fig1}(c) and (d) we picture the nested ellipsoids individually, so that also the inner ellipsoid (green) becomes visible. The ellipsoids touch along all principal directions in the BZ (S-R, T-R and U-R). Due to their nesting, $\xi_{\textbf{k},j}-\xi_{\textbf{k}+\textbf{q},j'}\sim 0$ in the denominator of Eq.~(\ref{eq:bare_susc}), which contributes to the peak around $\Gamma$ in the susceptibility, and thus to the enhancement of FSFs. 

Subsequently, we calculated the phonon dispersion and the \textit{e-ph} coupling in FeB$_4$ using DFPT. We show the phonon dispersion in Fig.~\ref{fig:fig2}(a). The highest phonon frequencies reach almost 120 meV, a considerably high value, due to the very light B atoms. This maximum frequency is even higher for FeB$_4$ than for other borides such as MgB$_2$ \cite{Bekaert2017,Bekaert2017b} and OsB$_2$ \cite{PhysRevB.94.144506}. It corroborates its extreme hardness, also mentioned in the introduction. Moreover, the difference in mass between Fe and B explains why their respective vibrational modes are well separated, as shown in the phonon density of states (PHDOS) in Fig.~\ref{fig:fig2}(b). In Fig.~\ref{fig:fig2}(c), we display the isotropic Eliashberg function, obtained from the full Eliashberg function as the double Fermi surface average $\alpha^2F(\omega)=\langle\langle\alpha^2F({\bf k\, k'},\omega)\rangle_{{\bf k}'_{\mathrm{F}}}\rangle_{{\bf k}_{\mathrm{F}}}$, and the resulting isotropic \textit{e-ph} coupling $\lambda(\omega)=2\int_0^\omega \mathrm{d}\omega'\omega'^{-1}\alpha^2F(\omega')$. The contributions of the two atomic species to the \textit{e-ph} coupling are comparable, in contrast to, e.g., MgB$_2$ (where B dominates) \cite{Choi2002,Bekaert2017,Bekaert2017b} and OsB$_2$ (where Os dominates) \cite{PhysRevB.94.144506}. The \textit{e-ph} coupling amounts in total to a very high value, $\lambda=1.13$ \footnote{We found that a high degree of interpolation of the \textit{ab initio} electron-phonon coupling was necessary to obtain a well-converged result. With a lesser degree of interpolation we find isotropic \textit{e-ph} coupling values similar to those of Ref.~\citenum{PhysRevLett.105.217003}, but with sufficient interpolation the \textit{e-ph} coupling increases to $\lambda_{\mathrm{ep}}=1.13$.}, much too high to corroborate the experimental $T_{\mathrm{c}}\sim 2.4$ K. This is the motivation for the following section, where we address FSFs in FeB$_4$. 
\begin{figure}[t]
\centering
\includegraphics[width=1\linewidth]{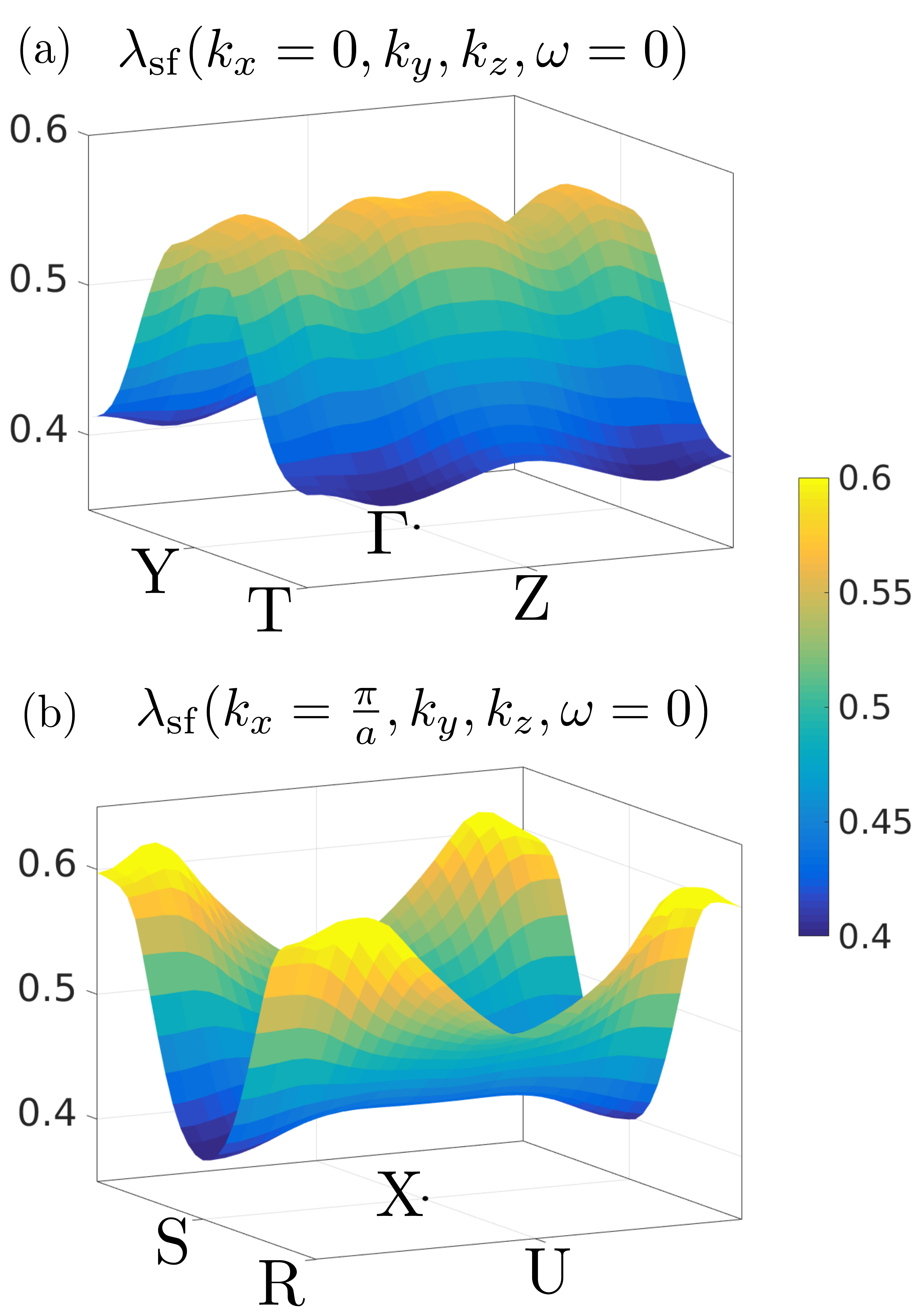}
\caption{(Color online) The coupling between electrons and FSFs in FeB$_4$,   $\lambda_{\mathrm{sf}}\left(\textbf{k}_{\mathrm{F}}, \omega_n\right)$, in the static limit ($\omega=0$), calculated using Eq.~(\ref{eq:coupling}), and convoluted with the Fermi surface to obtain the dependence on $\textbf{k}_{\mathrm{F}}$. (a) The coupling for $k_x=0$, (b) The coupling for $k_x=\frac{\pi}{a}$. The average over the whole Fermi surface is $\lambda_{\mathrm{sf}}=\langle \lambda_{\mathrm{sf}}\left(\textbf{k}_{\mathrm{F}}, \omega=0\right)\rangle_{\textbf{k}_{\mathrm{F}}}=0.55$.}
\label{fig:fig4}
\end{figure}
\section{Ferromagnetic spin fluctuations and their coupling to electrons}

In this section we demonstrate the occurrence of FSFs in FeB$_4$ from first principles, and we calculate their coupling to the electronic states. First, we calculated the bare susceptibility of FeB$_4$ at $E_\mathrm{F}$, from the band structure shown in Fig.~\ref{fig:fig1}, using Eq.~(\ref{eq:bare_susc}). The result is shown in Fig.~\ref{fig:fig3}(a), in the static limit, and for $q_x=0$. The peak in $\chi^0$ around $\Gamma$ indicates FSFs. It can be traced back to small-\textbf{q} intraband transitions, as well as to interband contributions of the nested ellipsoidal Fermi sheets, as discussed in the previous section. 

For small, though nonzero $q$, there are strong interband contributions to the peak, which ultimately vanish for $q \rightarrow 0$. The value of $\chi^0(0,0)$ therefore reduces in principle to the electronic DOS at $E_{\mathrm{F}}$. The numerical evaluation of Eq.~(\ref{eq:bare_susc}) depends, however, also on the broadening factor $\sigma$. We obtained that we needed a significant broadening of $\sigma=0.01$ Ha for a well-converged susceptibility. As such, $\chi^0(0,0)$ is artificially enhanced beyond the DOS at $E_{\mathrm{F}}$ [$N(E_{\mathrm{F}})=0.70$ states/(eV spin)]. The definition of $\mathcal{F}_{\mathrm{Fe}}$ in Sec.~\ref{Sec2}, entering in Eq.~(\ref{eq:susc_RPA}), nevertheless ensures that this enhancement cancels out completely in the RPA susceptibility. 

The susceptibility peaking at $\Gamma$ is a necessary condition for FSFs, but is not sufficient for a significant effect of these FSFs on superconductivity. The important other factor is the interaction strength, which is given by the Stoner parameter in the ferromagnetic case. We obtained the Stoner parameter by introducing nonzero magnetization into the material within a fixed spin moment (FSM) calculation, resulting in an energy shift between majority and minority spin states. The results of this calculation are depicted in Fig.~\ref{fig:fig3}(b), where the electronic density of states (DOS) in different states of magnetization is shown. In the case where $m=0$ ($m$ being the magnetization per Fe atom) we also depict the contributions of different atomic states. This DOS corresponds to the electronic structure shown in Fig.~\ref{fig:fig1}. The states at and near $E_{\mathrm{F}}$ are dominated by Fe-d and B-p character. We found $N_{\mathrm{Fe}}(E_{\mathrm{F}})=0.29$ states/(eV spin), compared to the total $N(E_{\mathrm{F}})=0.70$ states/(eV spin). For nonzero $m$, the energy shift between spin-up and spin-down bands can be clearly observed, and can be seen to increase with $m$ in Fig.~\ref{fig:fig3}(b). 

Extracting the shifts near $E_{\mathrm{F}}$, clearly visible in Fig.~\ref{fig:fig3}(b), for different values of $m$ we obtain Fig.~\ref{fig:fig3}(c). Within Stoner theory, this shift due to the magnetization obeys the linear relation $\Delta E=I m/ \mu_{\mathrm{B}}$. For low values of $m$, the linear relation is indeed obeyed, as shown in Fig.~\ref{fig:fig3}(c), whereby a fit yields $I=1.54$ eV. For higher values of $m$, the increase in $\Delta E$ weakens, as expected. Since we obtain that $N_{\mathrm{Fe}}(E_{\mathrm{F}})I=0.45<1$ (Stoner criterion), FeB$_4$ indeed does not have a ferromagnetic ground state. On the other hand, the Stoner parameter is certainly high enough to induce considerable FSFs. 

It is interesting to note the importance of restricting the interaction strength to the Fe states, mentioned already in Sec.~\ref{Sec2}. If the total DOS were used instead, FeB$_4$ would come out as marginally ferromagnetic according to $N(E_{\mathrm{F}})I=1.08>1$. This may be related to the tendency of DFT in local spin density approximation (LSDA) or generalized gradient approximation (GGA) to overestimate static magnetism \cite{PhysRevB.83.100505}. However, in our FSM calculations we found the magnetic moments to be completely localized on the Fe atoms. This provides us with the physical rationale for limiting the Stoner-type interaction to Fe states only. Thus, by avoiding to treat $I$ as a free parameter -- as in, e.g., Ref.~\citenum{PhysRevB.83.100505} -- we remain close to a fully \textit{ab initio} approach. 

The interaction strength plays a crucial role in the coupling of FSFs to electrons, according to Eqs.~(\ref{eq:susc_RPA}) and (\ref{eq:coupling}). We calculated $\lambda_{\mathrm{sf}}\left(\textbf{k}_{\mathrm{F}},\omega\right)$ for FeB$_4$ using these formulas, where we obtained the dependence on $\textbf{k}_{\mathrm{F}}$ by convolution with the Fermi surface. The result in the static limit, $\omega \rightarrow 0$, in particular $\lambda_{\mathrm{sf}}$ in the $\Gamma$-Y-T-Z plane ($k_x=0$), is shown in Fig.~\ref{fig:fig4}. It is observed that there is strong coupling to FSFs in the direction $\Gamma$-Y, since in this direction small $\textbf{q}$'s connect parts of the anisotropic Fermi sheet (red), evident from Fig.~\ref{fig:fig1}(b). In the other directions $\lambda_{\mathrm{sf}}$ drops significantly. In Fig.~\ref{fig:fig4}(b) we show $\lambda_{\mathrm{sf}}$ in the X-S-R-U plane ($k_x=\frac{\pi}{a}$), that cuts through the center of the nested ellipsoidal Fermi sheets. Here, the coupling $\lambda_{\mathrm{sf}}$ shows a broad peak around R, due to nesting  of these sheets with small \textbf{q}. It diminishes accordingly in all other directions. The total static coupling of FSFs to electrons, calculated as the Fermi surface average, amounts to $\lambda_{\mathrm{sf}}=\langle \lambda_{\mathrm{sf}}\left(\textbf{k}_{\mathrm{F}}, \omega=0\right)\rangle_{\textbf{k}_{\mathrm{F}}}=0.55$.

\begin{figure*}[t]
\centering
\includegraphics[width=0.94\linewidth]{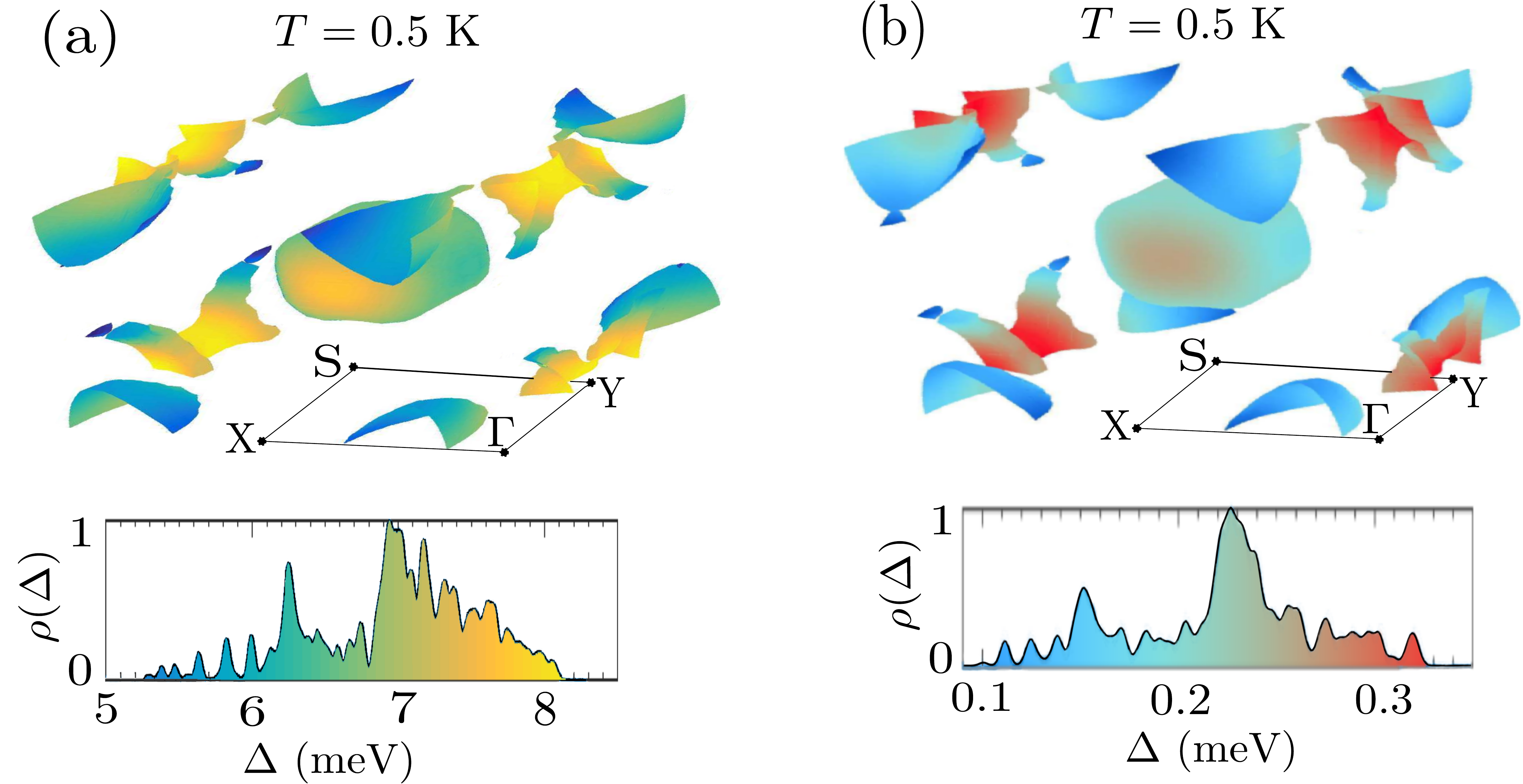}
\caption{(Color online) The superconducting gap spectrum, $\Delta(\textbf{k}_{\mathrm{F}},T)$, of FeB$_4$ on the Fermi surface at $T=0.5$ K, calculated using anisotropic Eliashberg theory with \textit{ab initio} input (and using $\mu^*=0.1$). (a) Gap spectrum obtained when only \textit{e-ph} coupling is taken into account. (b) Gap spectrum obtained when both \textit{e-ph} coupling and interaction with FSFs are taken into account. $\rho(\Delta)$ represents the distribution of the gap, showing a single, yet anisotropic gap in both cases. However, the superconducting gap is strongly depleted from the range $5-8$ meV to the range $0.1-0.3$ meV, under the influence of the FSFs.}
\label{fig:fig5}
\end{figure*}
\begin{figure}[t]
\centering
\includegraphics[width=1\linewidth]{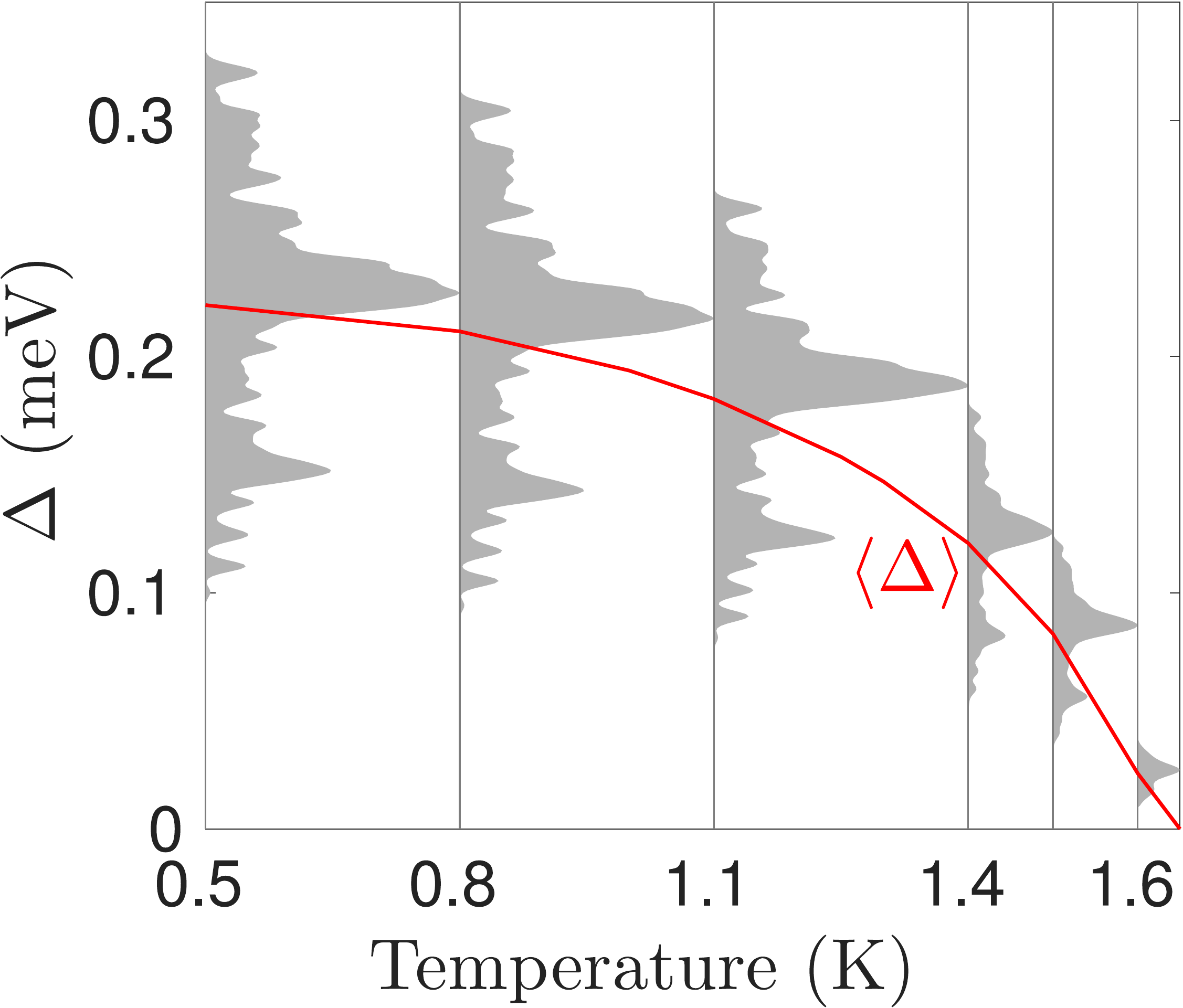}
\caption{(Color online) The superconducting gap distribution as a function of temperature, calculated using anisotropic Eliashberg theory with \textit{ab initio} input, including FSFs. The critical temperature can be seen to be $T_{\mathrm{c}}=1.65$ K. The red line represents the average value of the gap weighted with the distribution.}
\label{fig:fig6}
\end{figure}

\section{Superconducting properties}

Having established the \textit{e-ph} coupling and the coupling of electrons to FSFs in Secs.~III and IV, we can now study the competition between these interactions in relation to superconductivity. To this end, we solve the anisotropic Eliashberg equations including both the \textit{e-ph} coupling and the coupling to FSFs in the interaction kernel of the spin singlet channel. Here, we retain the full momentum and frequency dependence of the coupling. More information on how we solve the anisotropic Eliashberg equations numerically is provided in Appendix C.

We started by solving the anisotropic Eliashberg equations, taking into account just the \textit{e-ph} coupling, not yet the interaction with FSFs. We show the resulting superconducting gap spectrum $\Delta(\textbf{k}_{\mathrm{F}},T)$ on the Fermi surface, at low temperature ($T=0.5$ K) in Fig.~\ref{fig:fig5} (a). The gap spectrum consists of a single, anisotropic gap with large values for $\Delta$, ranging from 5 to 8 meV (again at $T=0.5$ K). The corresponding critical temperature obtained from solving the anisotropic Eliashberg equations for a range of temperatures is $T_{\mathrm{c}}=41$ K (using the standard value for the Coulomb pseudopotential, $\mu^*=0.1$). This value exceeds the experimental value ($T_{\mathrm{c}} \sim 2.4$ K) by more than an order of magnitude. 

When including the effect of FSFs, this changes drastically. In this case, the superconducting gap spectrum at low temperature ($T=0.5$ K) is shown in Fig.~\ref{fig:fig5} (b). $\Delta(\textbf{k}_{\mathrm{F}},T)$ presents again a single, anisotropic gap, but now in the range $\Delta(\textbf{k}_{\mathrm{F}},T)\sim 0.1-0.3$ meV. With such anisotropic, single-gap, FeB$_4$ is more similar to OsB$_2$ (an orthorhombic material, like FeB$_4$, also with three bands at $E_{\mathrm{F}}$) \cite{PhysRevB.94.144506} than to MgB$_2$ (a layered hexagonal material), which is a well-established two-gap boride superconductor \cite{0953-2048-16-2-305,Choi2002,PhysRevB.87.024505,PhysRevB.91.214519,PhysRevB.92.054516}. As the superconducting gap depletes rather uniformly under the influence of FSFs, as seen in Fig.~\ref{fig:fig5}, we conclude that the effect of FSFs is fairly isotropic in FeB$_4$. 

Subsequently, we again solved the anisotropic Eliashberg equations for a range of different temperatures, now taking into account both the \textit{e-ph} coupling and coupling to FSFs (using $\mu^*=0.1$ for the Coulomb pseudopotential). The resulting gap spectrum as a function of temperature is displayed in Fig.~\ref{fig:fig6}. The critical temperature we obtain is $T_{\mathrm{c}}=1.65$ K, in very good agreement with the experimental value $T_{\mathrm{c}} \sim 2.4$ K \footnote{In Ref.~\citenum{PhysRevLett.111.157002}, in Fig.~2 specifically, critical temperatures are obtained for FeB$_4$ with two different B isotopes. For $^{10}$B, $T_{\mathrm{c}}\sim 2.7$ K is obtained, while for $^{11}$B $T_{\mathrm{c}}\sim 2.4$ K. For our calculations we used a weighted average of these isotopes (mass of 10.81 atomic units), based on the natural occurrences of these isotopes. The resulting mass is closest to that of the heavier isotope, therefore our result can be compared to the experimental value $T_{\mathrm{c}}\sim 2.4$ K.}. FeB$_4$ is thus a superconductor with very strong \textit{e-ph} coupling, that in itself would lead to $T_{\mathrm{c}} =41$ K, which is however depleted to $T_{\mathrm{c}} \sim 2$ K due to FSFs. The very good agreement with the experimental value demonstrates that FSFs can be included in the anisotropic Eliashberg equations, to obtain a quantitatively accurate superconducting gap spectrum. To the best of our knowledge, our analysis is the first to report on this. 

In addition, we investigated the tendency for spin triplet pairing in FeB$_4$ due to FSFs. To this end, we adapted the Eliashberg kernels given in Sec.~II to the triplet case. The mass enhancement remains unaffected ($K^+=\lambda_{\mathrm{ep}}+\lambda_{\mathrm{sf}}$), while, on the other hand, FSFs become attractive in the spin triplet channel. Moreover, only longitudinal spin fluctuations contribute to spin triplet pairing \cite{Scalapino1999} (hence a factor $1/3$), so that the coupling kernel in the spin triplet channel amounts to $K^-=\lambda_{\mathrm{ep}}+\frac{\lambda_{\mathrm{sf}}}{3}$. Solving the anisotropic Eliashberg equations with these kernels, we did not obtain a gap function with symmetry $\Delta\left(\textbf{k}_{\mathrm{F}}\right)\rightarrow -\Delta\left(-\textbf{k}_{\mathrm{F}}\right)$, i.e., with an odd (ungerade) momentum dependence. Therefore, within our theoretical framework we can exclude the possibility of spin triplet pairing in FeB$_4$ \cite{Sigrist1991}. 

\section{Conclusion}

In summary, we presented an advanced approach to treat both lattice vibrations and ferromagnetic spin fluctuations in superconductors, entirely from first principles. Specifically, we extended the framework where the \textit{ab initio} calculated electron-phonon coupling ($\lambda_{\mathrm{ep}}$) is used to solve to anisotropic Eliashberg equations for the gap spectrum (which has been done for materials like MgB$_2$ \cite{Choi2002,PhysRevB.87.024505,PhysRevB.92.054516,Bekaert2017,Bekaert2017b} and OsB$_2$ \cite{PhysRevB.94.144506}). The first step is to calculate the bare susceptibility of the material from the electronic structure (in this work also calculated from first principles), specifically that near the Fermi level [cf.~Eq.~(\ref{eq:bare_susc})]. The next step is calculating the interaction strength of the ferromagnetic spin fluctuations. This can be achieved by means of the energy shift between minority and majority bands in the competing, ferromagnetic phase, which obeys $\Delta E=Im / \mu_{\mathrm{B}}$, where the Stoner parameter $I$ yields the interaction strength. Then, the RPA susceptibility can be calculated, yielding directly the coupling of ferromagnetic spin fluctuations with electrons $\lambda_{\mathrm{sf}}$ [using Eqs.~(\ref{eq:susc_RPA}) and (\ref{eq:coupling})]. For the spin singlet superconducting channel, the resulting total coupling is $\lambda_{\mathrm{ep}}-\lambda_{\mathrm{sf}}$, while for the spin triplet channel it amounts to $\lambda_{\mathrm{ep}}+\frac{\lambda_{\mathrm{sf}}}{3}$ \cite{Sigrist1991}. 

We have applied this new approach to the recently discovered Fe-based superconductor iron tetraboride (FeB$_4$) \cite{PhysRevLett.111.157002}, to resolve the large discrepancy between the predicted \cite{PhysRevLett.105.217003} and measured \cite{PhysRevLett.111.157002} critical temperature, and to learn more about its superconducting gap structure. We showed first that the Fermi surface has contributions from three different bands, resulting in two nested ellipsoids and an anisotropic sheet. This nesting at small $\textbf{q}$ is the main contribution to the peak in the calculated susceptibility of FeB$_4$, for small wavevectors, corresponding to ferromagnetic spin fluctuations. The Stoner parameter in FeB$_4$ is considerably high ($\sim 1.5$ eV) -- though not high enough for a ferromagnetic ground state. Accordingly, we found strong coupling of the spin fluctuations to the electronic states, in particular to the nested ellipsoids, with an average of $\langle \lambda_{\mathrm{sf}}\rangle_{\textbf{k}_{\mathrm{F}}}=0.55$ over the Fermi surface. This mediates a repulsive interaction between the electrons that is in direct competition with the strong, attractive interaction mediated by phonons (with Fermi surface average $\lambda_{\mathrm{ep}}=1.13$). By solving the anisotropic Eliashberg equations, we revealed that the spin fluctuations are able to reduce the critical temperature from a very high $T_{\mathrm{c}}=41$ K to $T_{\mathrm{c}}=1.7$ K, in very good agreement with the experimental value ($T_{\mathrm{c}}=2.4$ K \cite{PhysRevLett.111.157002}). In spite of this drastic effect on $T_{\mathrm{c}}$, we found that the distribution of the gap spectrum on the Fermi surface, namely a single anisotropic gap (similar to, e.g., OsB$_2$ \cite{PhysRevB.94.144506}), is largely unaltered. 

The excellent comparison between the results obtained with our new method and the corresponding experiment demonstrates the potential of an \textit{ab initio} approach to anisotropic Eliashberg theory in describing interactions that coexist and compete with the electron-phonon interaction. Although ferromagnetic spin fluctuations showed a primarily detrimental effect on superconductivity in FeB$_4$, we expect that our approach will lead to the detection and quantification of spin fluctuations in other materials with coexisting conventional and unconventional pairing mechanisms, resulting in nontrivial contributions to the superconducting gap spectrum and to the superconducting properties in general. 

\begin{acknowledgments}
\noindent This work was supported by TOPBOF-UAntwerp, Research Foundation Flanders (FWO), the Swedish Research Council (VR) and the R{\"o}ntgen-{\AA}ngstr{\"o}m Cluster. The computational resources and services used in this work were provided by the VSC (Flemish Supercomputer Center), funded by the Research Foundation Flanders (FWO) and the Flemish Government -- department EWI. Anisotropic Eliashberg theory calculations were  supported through the Swedish National Infrastructure for Computing (SNIC). 
\end{acknowledgments}

\section*{Appendix}

\begin{appendix}

\section{Crystal structure}
The oP10 FeB$_4$ phase (where o stands for othorhombic, P for primitive and 10 for the number of atoms in the unit cell) consists of the primitive orthorhombic space group Pnnm (No.~58). As can be found in the supplementary information of Ref.~\citenum{PhysRevLett.111.157002}, Fe occupies Wyckoff position $2a$, i.e., $\left(0, 0, 0\right)$ and $\left(\frac{1}{2}, \frac{1}{2}, \frac{1}{2}\right)$, and B Wyckoff position $4g$, i.e., $\left(\pm x, \pm y, 0\right)$ and $\left( \pm x + \frac{1}{2}, \mp y + \frac{1}{2}, \frac{1}{2}\right)$, where $x$ and $y$ are internal parameters. 

The results of our calculations are shown in Table \ref{tab:str_param}, and compared to the experimental values. It is observed that the deviations from the experimental values are all well below 1 \%. This very good agreement on the structural level propagates a high level of accuracy to all further calculations, of the electronic structure, phonons, spin fluctuations, and ultimately of the superconducting properties. 
\begin{table}[h!!!]
\centering
\begin{tabular}{|c|c|}
\hline
Experimental&Calculated\\ \hline \hline
$a=2.999$ \AA & 3.023 \AA~(+ 0.8 \%)\\ \hline
$a=4.579$ \AA & 4.552 \AA~(-0.6 \%)\\ \hline
$b=5.298$ \AA & 5.309 \AA~(+0.2 \%)\\ \hline
$x=0.249$& 0.247 (-0.8 \%)\\ \hline
$y=0.312$& 0.312 (+0.0 \%)\\ \hline
\end{tabular}
\caption{Comparison between experimental \cite{PhysRevLett.111.157002} (from room-temperature, single-crystal x-ray diffraction) and calculated structural parameters of FeB$_4$, obtained using PBE exchange-correlation. The relative deviations of the calculated parameters from the experimental ones are added between parentheses.}
\label{tab:str_param}
\end{table}

\section{Computational details on the \textit{ab initio} calculations}
Our density functional theory (DFT) calculations make use of the generalized gradient approximation (GGA), specifically of the Perdew-Burke-Ernzerhof (PBE) functional, implemented within a planewave basis in the ABINIT code \cite{Gonze20092582}. Electron-ion interactions are treated using norm-conserving pseudopotentials \cite{FUCHS199967}, taking into account Fe-3d$^7$4s$^1$ and B-2s$^2$2p$^1$ as valence electrons. The energy cutoff for the plane-wave basis was set to 60 Ha, to achieve convergence of the total energy below 1 meV per atom. To obtain a very accurate description of the Fermi surface, a dense $25 \times 15 \times 15$ $ \Gamma$-centered Monkhorst-Pack \textbf{k}-point grid was used. We use the notational convention established in Ref.~\citenum{Curtarolo} to denote the high-symmetry \textbf{k}-points. The optimized crystal structure was obtained using a conjugate-gradient algorithm so that forces on each atom were below 1 meV/\AA.

Density functional perturbation theory (DFPT) calculations of the phonon dispersion and the electron-phonon coupling coefficients were also carried out using ABINIT. Here, we employed a $25 \times 15 \times 15$ $\textbf{k}$-point grid for the electronic wavevectors and a $5 \times 3 \times 3$ $\textbf{q}$-point grid for the phonon wavevectors.

\section{Computational details on the Eliashberg calculations}

For the description of superconductivity on an \textit{ab initio} level, we need to solve self-consistently the coupled anisotropic Eliashberg equations with input from first-principles calculations \cite{Choi2002, PhysRevB.87.024505, PhysRevB.92.054516,PhysRevB.94.144506,PhysRevB.94.144506,Bekaert2017,Bekaert2017b}. For spin singlet superconductivity the coupled anisotropic Eliashberg equations assume the form,
\begin{eqnarray}\label{el1}
Z_{{\bf k},n}&=&1+\frac{\pi T}{\omega_n}\sum_{n'}\Bigl\langle K^+({\bf kk'},nn')\frac{\omega_{n'}}{\sqrt{\omega_{n'}^2 + \Delta^2_{{\bf k'},n'}}}\Bigl\rangle_{{\bf k}'}\\ \nonumber
\Delta_{{\bf k},n}&=&\pi T \sum_{n'}\Bigl\langle \left[K^-({\bf kk'},nn')- \mu^*(\omega_c)\right]\\\label{el2}
&\times&\frac{\Delta_{{\bf k'},n'}}{\sqrt{\omega_{n'}^2 + \Delta^2_{{\bf k}',n'}}}\Bigl\rangle_{{\bf k}'}/ Z_{{\bf k},n}
\end{eqnarray}
where $\bigl\langle\ldots\bigl\rangle_{{\bf k}'}=\sum_{{\bf k}'}\frac{\delta(\xi_{\bf k'})}{N_{\mathrm{F}}}(\ldots)$ denotes a Fermi surface average, $\xi_{\bf k}$ are electron energy dispersions, $N_{\mathrm{F}}$ is the density of states at the Fermi level, $T$ is temperature and $\omega_n=\pi T(2n+1)$ are fermion Matsubara frequencies. The momentum and frequency dependent functions $Z_{{\bf k},n}$ and $\Delta_{{\bf k},n}$ describe electron mass renormalization and even-frequency spin singlet superconductivity, respectively, and $\mu^*(\omega_{\mathrm{c}})$ is the Anderson-Morel Coulomb pseudopotential which comes with a cut-off $\omega_{\mathrm{c}}$. In the above, the following interaction kernels are used,
\begin{eqnarray}\label{kernel-singlet}
K^\pm({\bf kk'},nn')=\lambda_{\mathrm{ep}}({\bf kk'},nn') \pm \lambda_{\mathrm{sf}}({\bf kk'},nn')
\end{eqnarray}
that include the coupling of electrons to phonons, $\lambda_{\mathrm{ep}}({\bf kk'},nn')$, and spin fluctuations,  $\lambda_{\mathrm{sf}}({\bf kk'},nn')$. The momentum dependent electron-phonon coupling is
\begin{eqnarray}
\lambda_{\mathrm{ep}}({\bf kk'},nn')=\int_0^\infty d\Omega \, \alpha^2F({\bf k\, k'},\Omega) \frac{2\Omega}{\omega^2_m+\Omega^2} \,
\end{eqnarray}
with $\omega_m=\omega_n-\omega_{n'}$ and the momentum dependent Eliashberg function
\begin{eqnarray}
\alpha^2F({\bf k\, k'},\Omega)=N_F\sum_{\nu} |g^\nu_{\bf q}|^2 \delta(\Omega-\omega_{{\bf q}\nu}),
\end{eqnarray}
where $g^\nu_{\bf q}$ and $\omega_{{\bf q}\nu}$ are the phonon branch-resolved electron-phonon scattering matrix elements and phonon frequencies, respectively. From the above, one can obtain the  isotropic Eliashberg function as 
\begin{eqnarray}
\alpha^2F(\Omega) = \langle\langle\alpha^2F({\bf k\, k'},\Omega)\rangle_{{\bf k}}\rangle_{{\bf k}'}.
\end{eqnarray}
Similar equations apply also for the electron-spin fluctuation coupling,
\begin{eqnarray}\label{e-spinkernel}
\lambda_{\mathrm{sf}}({\bf q},m)=\frac{3N_{\mathrm{F}}}{2\pi}\int_0^\infty d\omega \,  I^2\chi''({\bf q},\omega) \frac{2\omega}{\omega^2_m+\omega^2} \,
\end{eqnarray}
where $\chi''({\bf q},\omega)$ is the imaginary part of the RPA susceptibility.

For spin triplet superconductivity in the unitary limit \cite{Sigrist1991}, Equations (\ref{el1}) and (\ref{el2}) still apply except from the fact that the kernel $K^-({\bf kk'},nn')$ needs to be substituted by 
\begin{eqnarray}\label{kernel-triplet}
K_{\mathrm{t}}({\bf kk'},nn')=\lambda_{\mathrm{ep}}({\bf kk'},nn') + \frac{1}{3}\lambda_{\mathrm{sf}}({\bf kk'},nn')
\end{eqnarray}

It is worth noting that even in the presence of spin-orbit coupling, inversion and time-reversal symmetries and the combination thereof, still guarantee Kramers degeneracy throughout the whole Brillouin zone, so that one can  work in a pseudospin space where labelling the superconducting state as singlet or triplet is possible. Moreover, due to inversion symmetry, singlet and triplet superconducting components cannot mix, so that we can look for different solutions (singlet or triplet) of the Eliashberg equations separately.

The numerical solution  of equations (\ref{el1}-\ref{el2}) along with an efficient calculation procedure  of the coupling in equation (\ref{e-spinkernel}) were implemented in the Uppsala Superconductivity Code (UppSC) \cite{PhysRevB.92.054516}. Using UppSC, the coupled equations (\ref{el1}-\ref{el2}), supplemented by the electron and phonon band structure and the electron-phonon and electron-(para)magnon coupling, calculated by first principles, were solved self-consistently in Matsubara space within a strict convergence criterion of $\frac{x_n-x_{n-1}}{x_n}<10^{-6}$ and with up to 1000 iteration cycles allowed. In all the calculations presented here we set $\mu^*(\omega_c)=0.1$ for the Coulomb pseudopotential with a sufficient value of $\omega_c$ to ensure that the results are well converged.

\end{appendix}

\bibliography{biblio}% Produces the bibliography via BibTeX.

\end{document}